# Automated, Cross-Layer Root Cause Analysis of 5G Video-Conferencing Quality Degradation

## Fan Yi
Princeton University
fanyi@princeton.edu

## Haoran Wan
Princeton University
haoran.w@princeton.edu

## Kyle Jamieson
Princeton University
kylej@princeton.edu

## Oliver Michel
Princeton University
omichel@princeton.edu

## Abstract

5G wireless networks are complex, leveraging layers of scheduling, retransmission, and adaptation mechanisms to maximize their efficiency. But these mechanisms interact to produce significant fluctuations in uplink and downlink capacity and latency. This markedly impacts the performance of real-time applications, such as video conferencing, which are particularly sensitive to such fluctuations, resulting in lag, stuttering, distorted audio, and low video quality. This paper presents a cross-layer view of 5G networks and their impact on and interaction with video-conferencing applications. We conduct novel, detailed measurements of both private CBRS and commercial carrier cellular network dynamics, capturing physical- and link-layer events and correlating them with their effects at the network and transport layers, and the video-conferencing application itself. Our two datasets comprise days of low-rate campus-wide Zoom telemetry data, and hours of high-rate, correlated WebRTC-network-5G telemetry data. Based on these data, we trace performance anomalies back to root causes, identifying 24 previously unknown causal event chains that degrade 5G video conferencing. Armed with this knowledge, we build **Domino**, a tool that automates this process and is user-extensible to future wireless networks and interactive applications.

## 1 Introduction

Real-time communication applications, such as interactive video conferencing and telephony, are ubiquitous in today's digital landscape, enabling seamless communication and collaboration across distances. Over the past years, their use cases have expanded massively from video conferencing for business or education; today, applications like FaceTime and WhatsApp are heavily used from cellular devices and commonly replace regular phone calls [9, 21, 24, 27, 32]. Common across these *real-time communication* (RTC) applications is a sensitivity to high and fluctuating latency and a demand for stable throughput [9, 12, 30, 32, 35, 36].

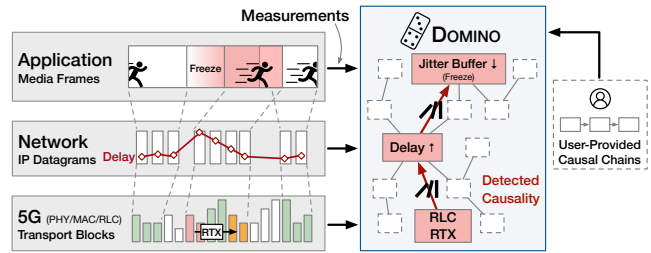

**Fig. 1— Domino detects causality chains leading to performance degradation from cross-layer measurements.**

In contrast to traditional applications such as web browsing and on-demand video streaming, RTC applications have several unique characteristics. First, they rely on uplink bandwidth and link quality since they involve bidirectional communication. Second, participants usually connect to the Internet from various locations, including home, office, or public spaces, using different (often wireless) access network technologies such as cellular or Wi-Fi networks. Last, where many applications can tolerate some delay stemming from retransmissions, RTC requires minimal packet loss and low latency to maintain a satisfactory user experience.

Wireless networks are inherently complex, relying on scheduling, duplexing, retransmission, and multiple-access mechanisms that introduce significant fluctuations in available capacity and latency [13, 28, 30, 33, 35, 36] which are particularly pronounced in the 5G uplink (§2.1). RTC applications have mechanisms to adapt to changing network conditions through congestion-control algorithms (*e.g.*, Google Congestion Control in WebRTC [5, 8]) that adjust media quality and transmission rate based on live feedback from receivers. Furthermore, applications leverage jitter buffers to smooth out variations in packet arrival times, ensuring a more consistent user experience.

However, while these mechanisms are effective in many scenarios, they are not sufficient for modern 5G cellular networks. Our real-world, 500-day, campus-wide measurement dataset demonstrates how video conferencing performs consistently worse over cellular networks compared to wired networks and even Wi-Fi (§2.2). The impact on user quality of experience (QoE) is significant: even transient periods of increased delay can cause noticeable lag in conversations, video stuttering, and distorted audio, heavily degrading interactivity. Additionally, rapid fluctuations in link conditions can mislead congestion-control algorithms, causing them to overreact and unnecessarily throttle transmission rates, and cause the jitter buffer to increase lag and hinder interactivity [11]. This excessive throttling and slow recovery, combined with temporary latency spikes, consistently leads to poor QoE for users.

While the problem is well known [13, 17, 35, 36], the underlying causes of these performance issues remain poorly understood. In particular, it is crucial to move beyond anecdotal evidence and systematically investigate the underlying causes. We need to understand how different types of 5G network effects—such as latency spikes, bandwidth fluctuations, and retransmissions—impact the performance of real-time communication applications. This requires a comprehensive and methodical approach to measurement, correlation, and analysis, capturing the full scope of cellular network dynamics. By examining these effects across all layers, from the intricacies of 5G to the high-level decision-making processes in the application layer, we can develop a deeper understanding of the challenges and potential solutions for improving real-time application performance.

This paper presents **Domino**, a measurement framework and automated, cross-layer causal-chain detection tool that scrutinizes RTC application performance by looking across all layers of the stack, from the cellular physical and link layers, through the network and transport layers, and ending at the RTC application itself (§4). This enables a previously unprecedented cross-layer view of cellular networks and their impact on video conferencing applications. To build this framework, we have conducted first-of-kind, detailed measurements of WebRTC in cellular networks, capturing physical- and link-layer events and correlating them with their effects at the transport and application layers (Sections 3 and 4.1). WebRTC is an open-source framework to build RTC applications that is implemented in all major browsers and widely used in video conferencing applications (VCA), including Google Meet and Microsoft Teams. Our study encompasses both private 5G CBRS small cells and commercial 5G networks operated by a major carrier, allowing for a comprehensive analysis across different deployment scenarios.

From our measurements and a detailed analysis of 5G (§5) across six datasets (Table 1), we identify a series of precise causal relationships across layers, originating from events in the physical, medium access control, and radio link control layers of the cellular Radio Access Network (RAN), leading to delays in the network layer, impacting the operation of the

RTC transport layer, and finally terminating at multiple possible user-visible impacts at the application layer. By bridging the gaps between low-level network behavior and RTC application performance, Domino yields valuable insights into the challenges faced by video conferencing applications in cellular environments. In our commercial 5G dataset, Domino identifies an average of approximately five video quality degradation events per video session per minute, attributing these degradations primarily to cross-traffic (28%), retransmissions (42%), and poor-quality channels (12%). In our private 5G dataset, uplink scheduling delays (36%) and poor quality channels (37%) dominate the causes.

Using these findings, we develop the Domino automated analysis tool that automates the identification of the root causes of performance issues in RTC applications, given cross-layer trace data, which network operators can provide on a continuous, near real-time basis. The Domino tool models a graph of many (often overlapping) causal relationships across all layers and applies a search algorithm to traces of network events and application performance metrics to find individual paths through this graph (i.e., causal chains) that lead to the most likely root cause of performance issues. Domino exposes a simple configuration API that allows adding new causal relationships through a simple text configuration file. Domino enables network operators, application developers, and researchers to understand and address performance issues in RTC applications in modern cellular networks. The high-level architecture of Domino is depicted in Fig. 1. We will make Domino and our high rate 5G datasets available to the community upon publication of this work.

## 2 Motivation: 5G Considered Harmful

Video-conferencing applications (VCAs) are particularly sensitive to network conditions: frequent short-term fluctuations in available capacity and latency can severely degrade quality of experience (QoE) for users [35]. This degradation typically manifests in one or more of the following ways: First, rapidly changing network delay can lead the congestion controller to misinterpret the network state, causing it to either over- or under-utilize the available bandwidth. Second, in the presence of a short jitter buffer, the application may not be able to smooth out the network jitter, leading to frame-rate drops, audio stuttering or distortion, and even video freezes. Third, a long jitter buffer may lead to smooth and uninterrupted playback at the cost of increased end-to-end latency which hinders interactivity [11, 12, 18]. This delay is often referred to as *mouth-to-ear delay*.

### 2.1 WebRTC Cellular Performance

To motivate our work, we present an experiment where we compare the performance of a WebRTC video-conferencing session over a commercial 5G cellular network and a wired connection (both sessions were conducted in series). One client is in our lab while the other client runs on a public cloud server approximately 150 miles from our campus. Details about the experimental setup (which we use for all our experiments) can be found in Section 3.

| Dataset | Type | Configuration | | | Duration | Time | DCI | Event Rate (per min.) | | | Variables |
| | | Frequency | Bandwidth | Duplex. | | | | gNB | Pkt. | WebRTC | |
|---|---|---|---|---|---|---|---|---|---|---|---|
| T-Mobile 1 | Public | 622.85 MHz | 15 MHz | FDD | 120 min. | Day | 37941 | 0 | 98241 | 12976 | 227 |
| | | | | | 30 min. | Night | 36469 | 0 | 120051 | 13154 | 227 |
| T-Mobile 2 | Public | 2506.95 MHz | 100 MHz | TDD | 120 min. | Day | 14052 | 0 | 119979 | 8664 | 227 |
| | | | | | 30 min. | Night | 13731 | 0 | 129811 | 9759 | 227 |
| Amarisoft | Private | 3547.20 MHz | 20 MHz | TDD | 60 min. | Day | 29094 | 29094 | 96691 | 12143 | 247 |
| Mosolabs | Private | 3630.72 MHz | 20 MHz | TDD | 60 min. | Day | 31683 | 0 | 132215 | 9542 | 227 |
| Zoom API | | (Organization-wide API data) | | | 500 days | Day+Night | | | 1 | | 62 |

**Table 1— Overview of datasets used in this study: collection duration, event rates, and variable record sizes.**

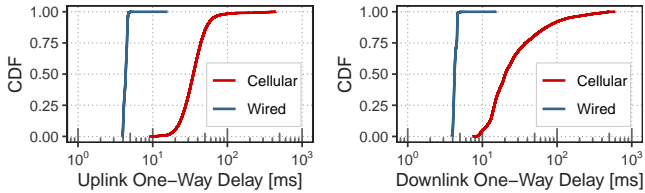

**Fig. 2— 5G v. wired network one-way packet delay: 5G inflates median delay by 1-2 orders of magnitude.**

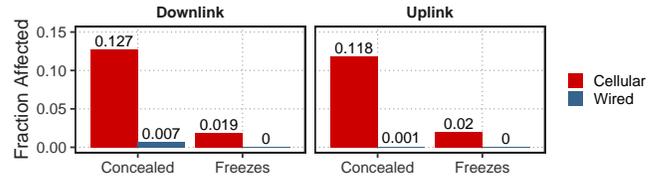

**Fig. 4— WebRTC concealed audio samples and freezes: cellular v. wired network.**

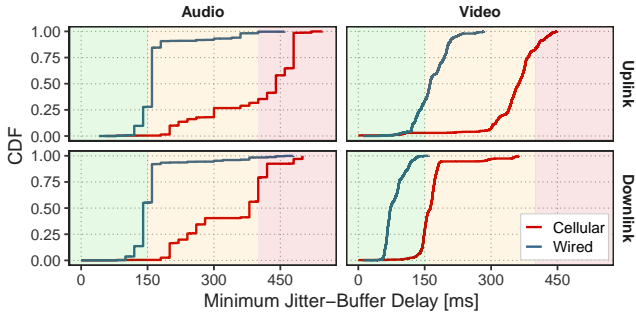

**Fig. 3— 5G v. wired network jitter-buffer delay: 5G more often impacts interactivity (see thresholds indicated).**

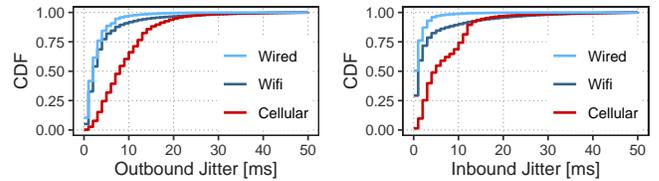

**Fig. 5— Campus Zoom Dataset: network jitter.**

**5G's end-to-end delay impact.** Fig. 2 shows the one-way network packet delay for both the uplink and downlink of the WebRTC session over 5G and a wired connection. We observe that 5G network delay dominates the wired network delay, and also exhibits high variance, with 99th percentile delays of 352 and 381 ms for the uplink and downlink, respectively, which has a direct impact on the size of the jitter buffer at the receiver [11]. Fig. 3 shows the *jitter-buffer delay* (the time a video frame or audio sample is held in the jitter buffer before being released for playback) for both uplink and downlink. The sum of the packet one-way delay plus the jitter-buffer delay represents a lower bound on the mouth-to-ear delay, a critical QoE metric. Like network delay, this delay is significantly higher for 5G network than for wired. The ITU-T states that mouth-to-ear delays of more than 150 ms can impact interactivity (orange area) and that delays of more than 400 ms is considered unacceptable (red area) [18].

**5G's impact on playback quality.** WebRTC provides detailed metrics about the playback quality of media streams, among them the time of a video stream in a frozen state, as well as the number of *concealed* audio samples that were not played back but instead replaced by a synthetically-generated sample [2]. Fig. 4 shows the fraction of concealed audio samples and the total freeze duration for both the uplink and downlink of the WebRTC session over 5G and a wired connection. In the five-minute experiment, approximately 12% of audio samples were concealed and the video stream was frozen for a total of six seconds, while the wired experiment showed few concealed audio samples and no video freezes.

Taken together, 5G networks significantly impact both the experienced delay and the playback quality of VCAs.

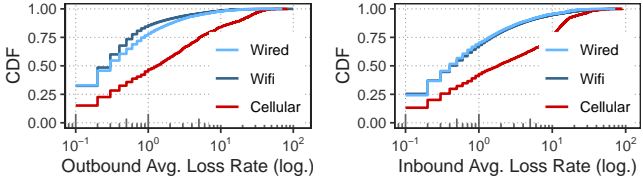

**Fig. 6— Campus Zoom Dataset: packet loss rate.**

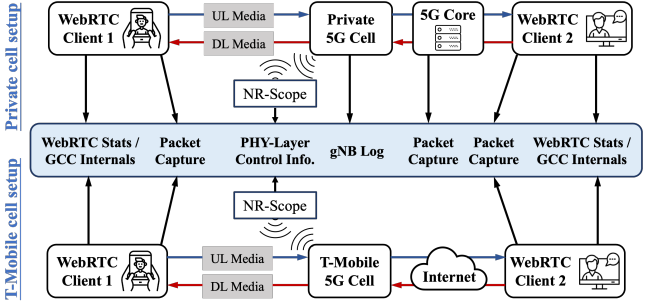

**Fig. 7— Experimental setup for two-party WebRTC measurements conducted over T-Mobile commercial cells (bottom) and private 5G cells (top).**

## 2.2 Campus-Wide Zoom Network Metrics

To demonstrate that 5G performance degradation is pervasive and not limited to the previous experiment, we analyze Zoom quality of service (QoS) metrics from all meetings conducted on our campus over a time period of one week in February 2023. Zoom enterprise customers can access various QoS metrics of all meetings that involved at least one participant who was signed into Zoom using an account associated with our institution. Our metrics report through which type of access network (wired, Wi-Fi, or cellular) the participant was connected to Zoom [37] and the network conditions each participant experiences at one-minute intervals. The dataset comprises 409 days of Wi-Fi, 86 days of wired, and 165 hours of cellular network data.

Comparing meeting quality metrics from this data set can be misleading as, for example, Zoom often reduces frame rate and resolution when participants use Zoom's gallery view mode [24]. Figure 5 shows the network jitter per access network type (Wired, Wi-Fi, or cellular). Note that cellular here refers to any cellular generation (i.e., 3G, 4G, or 5G). *Inbound* refers to the statistics received by clients (i.e., downlink) while *outbound* refers to the send side (i.e., uplink). We observe that jitter is consistently higher for cellular networks than for wired and Wi-Fi networks. We plot the packet loss rate in Fig. 6: the cellular network shows a significantly higher packet loss rate than the wired or Wi-Fi networks.

In summary, cellular networks consistently show higher network jitter and packet loss than wired and Wi-Fi networks.

## 3 Longitudinal 5G WebRTC Performance

Motivated by the challenges 5G networks pose to VCA performance, this section longitudinally measures VCA quality degradation through a comprehensive experimental study. We begin by outlining our experimental setup and data collection from multiple operational 5G cells. Following this, we characterize the measured WebRTC performance issues.

**Experimental setup.** To systematically study the impact of 5G network dynamics on VCA quality, we conduct a series of experiments on two-party meetings, as depicted in Fig. 7. Each experiment consists of a 30-minute WebRTC call with one client connected to a 5G network and the other to a wired network; all host clocks were synchronized using NTP for accurate temporal data correlation. We perform 14 such calls, distributed across four distinct 5G cells: two commercial T-Mobile 5G networks and two Private 5G networks (Amarisoft [3] and Mososlabs [25]): Table 1 summarizes the specification of each. To minimize variability due to video content we inject a prerecorded video file via a virtual camera device for both clients in all calls. For measurements on commercial cells (Fig. 7 lower), the wired client was hosted on a Google Cloud Platform (GCP) [15] server. In the private cell counterpart (Fig. 7 upper), the wired client ran on a local server in the same subnet as the Private 5G core.

To precisely capture event chains, we collect high-rate information (table 1) across layers 2 to 7:

**1) 5G Protocol Stack.** We use NR-Scope [33] to gather sub-millisecond resolution measurements from the 5G PHY/MAC layers, including traffic scheduling information and retransmission events. In the private-cell measurements, we also collect base station (gNB) logs which provide insights into Radio Link Control (RLC) layer buffer status and retransmissions, and the Radio Resource Control (RRC) layer state.

**2) Network Layer.** We collect packet traces at both clients. In the private cells, we added an additional capture point within the 5G core to more accurately isolate RAN delay.

**3) Application Layer.** We use a custom WebRTC client built on top of libwebrtc [16] and written in C++. This custom client allows us to gather quality statistics more frequently (every 50 ms as opposed to every second in the JavaScript API). These statistics include frame rate, resolution, freeze statistics, and jitter-buffer delay, among others [2]. Moreover, we collect internal state from GCC, including delay variation, perceived network state, target bitrate, and pushback rate. To the best of our knowledge, this is the first work to instrument WebRTC to this level which gives us detailed insight into the exact behavior of GCC in challenging network conditions.

**Experimental results.** Our analysis of WebRTC performance across the four distinct 5G cellular environments includes metrics for both uplink (UL, blue curves) and downlink (DL, red curves) in Fig. 8. Our analysis of one-way packet delay (Fig. 8a-d) reveals that UL streams consistently exhibit higher median delays than their DL counterparts across all four cells. This general trend is primarily attributed to the overhead of 5G UL scheduling mechanisms (§5.2.1). A notable deviation from this pattern is observed in the T-Mobile 15 MHz FDD cell (Fig. 8b), where the DL stream displays a significantly longer delay tail than the UL. This indicates a higher incidence of severe delay anomalies specifically in the DL, a phenomenon we associate with the characteristics of this heavily utilized commercial cell. Prevalent asymmetric traffic patterns, where users generate significantly more DL cross traffic, contrasting with the symmetric WebRTC load, contribute to this long-tail delay distribution (§5.1.2).

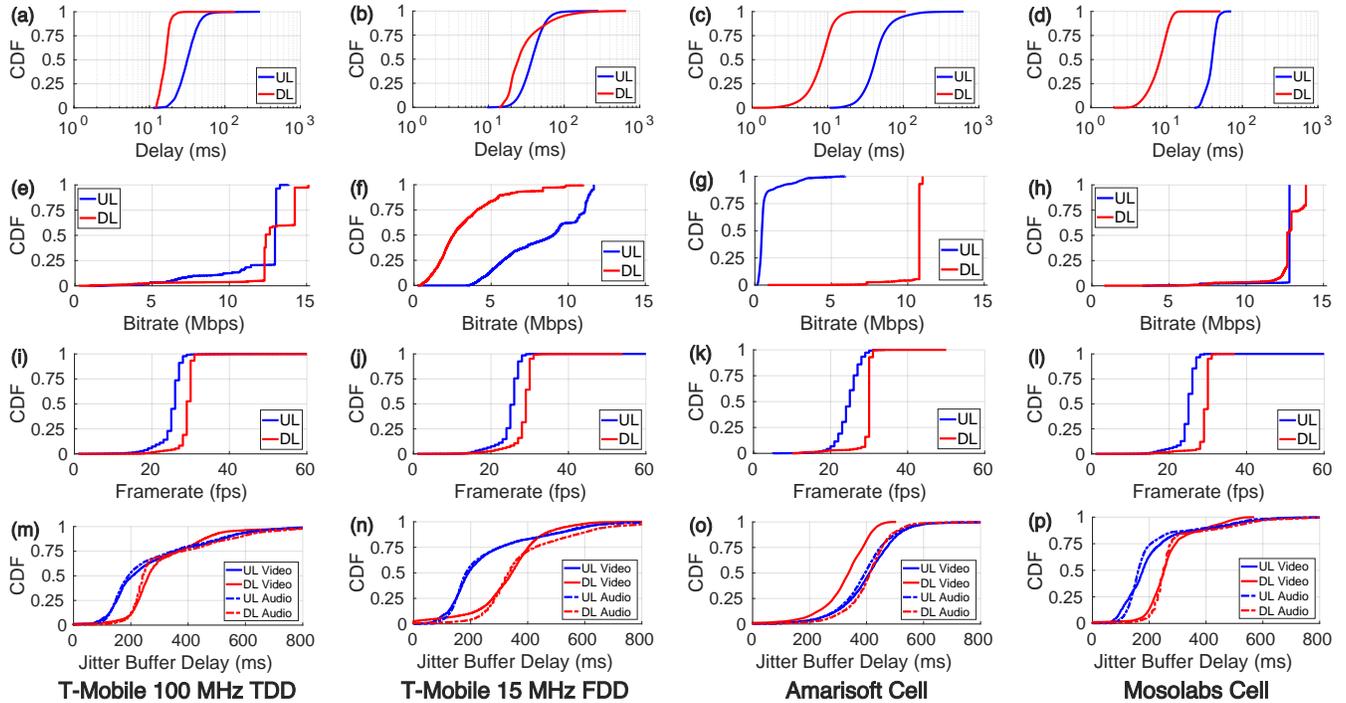

**Fig. 8— WebRTC performance metrics across four 5G cells: (a)-(d) one-way delay between two WebRTC clients, (e)-(h) target bitrate, (i)-(l) frame rate at the receiver side, (m)-(p) jitter-buffer delay at the receiver side.**

These frequent DL cross traffic events in the T-Mobile 15 MHz FDD cell also significantly influence Google Congestion Control (GCC) decisions on achievable bitrates. As illustrated in Fig. 8f, the DL target bitrate for this cell is considerably lower than its UL counterpart. This differs from the other three cells, where DL target bitrates generally exceed UL bitrates (Fig. 8e,g,h). Separately, the Amarisoft cell exhibits a markedly lower UL bitrate compared to its DL (Fig. 8g). This substantial gap primarily results from persistent poor UL channel conditions, coupled with the cell's conservative UL MCS selection strategy (§5.1.1).

Turning to frame rates, DL streams typically achieve higher frame rates than UL streams across all cells (Fig. 8i-l). Finally,

examining jitter buffer delay CDFs (Fig. 8m-p), median values typically range between 200 to 250 ms across most streams. However, reflecting the previously discussed network challenges, the DL stream in the T-Mobile 15 MHz FDD cell and the UL stream in the Amarisoft cell exhibit relatively higher jitter buffer delays. These increases stem from the impact of DL cross traffic in the former case and persistent poor UL channel conditions in the latter.

## 4 Domino: Tracing VCA QoE Impairments

Given these VCA consequences, we now describe **Domino**, an extensible tool designed for automated detection and statistical analysis of causal chains that trace a user-reconfigurable directed acyclic graph through cross-layer data from each root cause to each consequence.

### 4.1 WebRTC Anomaly Causation Overview

Leveraging our comprehensive cross-layer measurements, we have traced each WebRTC quality degradation to specific root causes within the 5G protocol stack. Given GCC's sensitivity to network delay, our causal analysis, illustrated in Fig. 9, focuses on six principal 5G causes (yellow blocks) that induce delay increases (purple blocks) and three consequences in WebRTC (red blocks) stemming from such delay.

Two major factors affecting physical-layer capacity are channel condition dynamics (§5.1.1) and 5G cross traffic

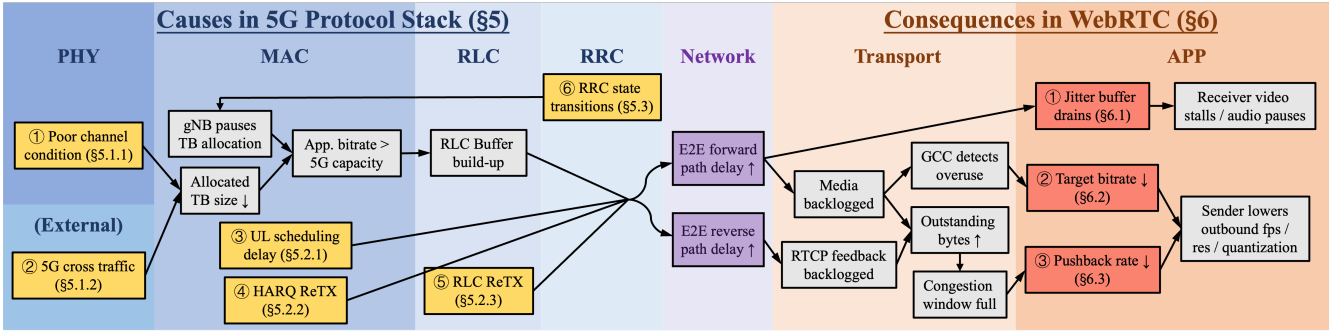

**Fig. 9— Causality graph of WebRTC quality degradations, illustrating six root causes (yellow blocks) across different layers of the 5G protocol stack and three consequences (red blocks) at the WebRTC application layer.**

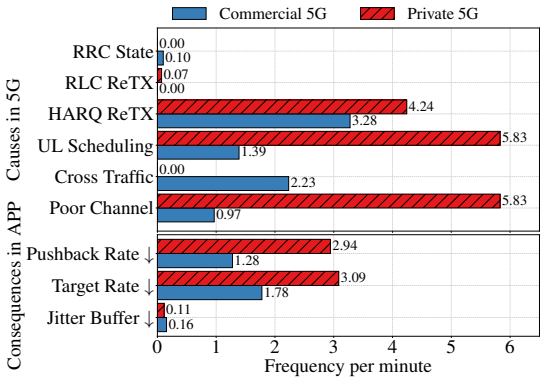

**Fig. 10— 5G network cause and VCA consequence absolute occurrence frequency.**

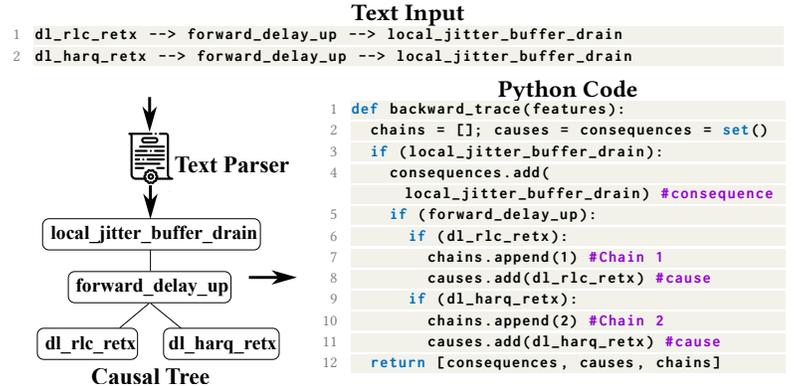

**Fig. 11— Domino generates Python code from text input; the code is translated to human-readable pseudo-code here.**

(§5.1.2). Both can lead to a drop in the achievable physical-layer data rate, subsequently causing 5G RLC layer buffer build-up and thus increased one-way delay.

5G protocol timing and reliability mechanisms (§5.2) introduce further delay inflation. These mechanisms give rise to three distinct causes: UL scheduling delay (§5.2.1), H-ARQ Retransmission (ReTX) delay (§5.2.2), and RLC ReTX delay (§5.2.3). Additionally, we identify RRC state transition delays (§5.3) as another significant source of delay inflation.

These six 5G-induced causes can trigger three primary consequences at the WebRTC application layer. Jitter buffer draining (§6.1) leads to receiver-side video playback stalls and audio pauses. GCC estimated target bitrate reduction (§6.2) and GCC pushback rate (§6.3) are both WebRTC sender-side mechanisms to proactively lower the outbound video quality. All three significantly impair QoE.

### 4.2 Automated Causal Chain Detection

Based on the causal relationships depicted in Fig. 9, we define 24 potential causal chains by analyzing all combinations of the causes and consequences interconnected through the paths in this directed acyclic graph.

**Event detection.** After gathering the time-series data from multiple sources, Domino maintains a sliding *window* of length $W = 5$ s. Within the window, Domino determines whether one or more events in Fig. 9 happen by testing all data within that window (enumerated by index $i$) against a series of *event conditions*.

For example, Domino identifies consequence ① (jitter buffer drains) when it finds the buffer length decreases to 0 ms: $\exists i \in [0, W]$, jitter_buffer_delay$[i] = 0$ ms. Domino detects cause ② (cross traffic) when the PRBs allocated for other users (PRB$_j$, $j \in [1, n]$) exceed 20% of the PRBs of the target client (PRB$_0$): $\sum_{i \in W} \sum_{j=1}^{n} \text{PRB}_j[i] > 0.2 \sum_{i \in W} \text{PRB}_0[i]$.

For each window, Domino generates a 36-dimension feature vector, and uses it for causal chain detection.[1] After processing each window, Domino moves the window edge forward by a *step length* $\Delta t = 0.5$ s throughout the vectorized data, and re-runs the event and further causal chain detection within the new window.

**Event and chain statistics.** We present the absolute (unconditional) occurrence frequency of individual 5G causes and WebRTC consequences in Fig. 10, distinguishing between commercial (blue) and private (red) 5G cells. Complementing this, Table 2 provides the conditional probability of each identified 5G cause being associated with a specific WebRTC consequence event. Table 2 reveals that UL scheduling delays and HARQ ReTXs are quite prevalent across both commercial and Private 5G networks. However, it is noteworthy that while these two mechanisms frequently inflate one-way delay, they are often not the primary contributor to severe quality degradation. Our findings also indicate that our Private 5G testbeds experience poor channel conditions more frequently, largely due to persistent UL channel issues in the Amarisoft cell. The absence of RLC ReTX detections in commercial cells is because their RLC-layer information is unavailable. Furthermore, disruptive RRC state transitions during active sessions were uniquely observed in the T-Mobile 15 MHz FDD cell.

---

[1]See Appendix D for the complete list of event conditions.

| | Poor Channel | | Cross Traffic | | UL Scheduling | | HARQ ReTX | | RLC ReTX | | RRC State | | Unknown | |
|---|---|---|---|---|---|---|---|---|---|---|---|---|---|---|
| **Jitter Buffer Drains** | 9.7% | 22% | 26% | 0% | 23% | 22% | 42% | 22% | 0% | 12% | 0% | 0% | 0% | 22% |
| **Target Bitrate ↓** | 9.9% | 43% | 32% | 0% | 15% | 43% | 42% | 12% | 0% | 0.42% | 0.80% | 0% | 0.27% | 1.9% |
| **Pushback Rate ↓** | 13% | 31% | 23% | 0% | 19% | 31% | 38% | 37% | 0% | 0.31% | 5.7% | 0% | 1.1% | 0.15% |

**Table 2— Conditional probability of causes given the consequence under commercial cells (blue), and private cells (red), where rows represent the consequences and columns represent the causes.**

Regarding consequence occurrence frequency (Fig. 10, bottom subplot), jitter buffer draining events are less frequent than GCC-initiated bitrate reductions or pushback rate drops. We attribute this to GCC's proactive rate control mechanisms, which react to network congestion by reducing sending rates, thereby preemptively preventing jitter buffer draining.

**Extensibility of Domino.** A key design principle of Domino is its extensibility, facilitating adaptation to new causal chain detection tasks. As illustrated in Fig. 11, Domino generates Python detection code directly from a user's textual causal chain definition. Domino parses the text input, mapping its constituent elements (causes, intermediate events, consequences) to corresponding feature vector entries. Then it constructs an internal graph representation of the specified causal chains, and finally this structured representation is used to automatically generate executable Python code that identifies and reports instances of the defined causal chain, including its cause, intermediate nodes, and consequence.

This extensibility allows network designers to readily incorporate other data features, such as other metrics from NR-Scope [33] or additional WebRTC statistics, and implement detection for novel causal chains simply by providing new text-based definitions. We believe Domino's approach can also be effectively generalized to other anomaly detection tasks that benefit from structured causal chain analysis.

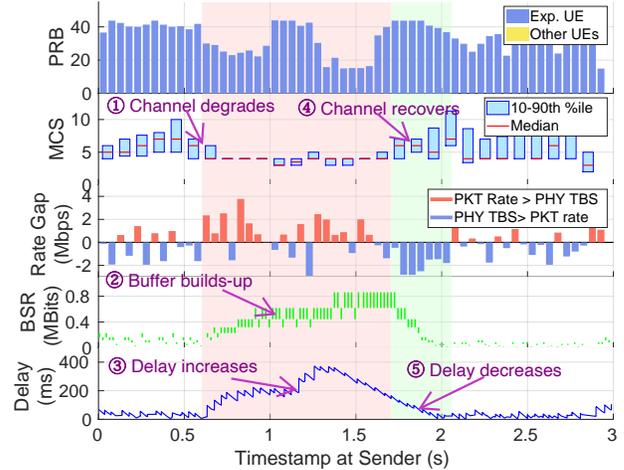

**Fig. 12— 5G channel condition dynamics cause RLC buffer build-up and delay increase.**

## 5  5G Causes of VCA Quality Degradation

We now analyze the mechanisms underlying the causal chains presented in the previous section, analyzing how events in the 5G stack cause capacity and latency fluctuations at higher layers that impact VCA quality. Then in §6 we will make the connection to VCA quality consequences.

### 5.1  5G Radio Resource Variability

5G New Radio manages radio resources using a time-frequency grid (Fig. 15a). The fundamental unit of resource allocation is the *Physical Resource Block* (PRB). The radio resources allocated to one *User Equipment* (UE) over one time slot constitute a *Transport Block* (TB). The *Transport Block Size* (TBS) depends on the number of allocated PRBs and the wireless physical-layer bit rate. This bit rate is primarily determined by the *Modulation and Coding Scheme* (MCS), which is selected based on the UE's wireless channel conditions. The achievable TBS for a UE can be highly variable, influenced by fluctuations in channel quality (due to mobility, fading, or interference) and cross traffic.

*5.1.1 Impact of Channel Condition Dynamics.* Dynamic and poor channel quality necessitates the use of a lower and more robust MCS, which directly reduces the TBS for a given number of PRBs. This happens frequently in 5G networks, especially in urban environments with high interference and fading. Fig. 12 illustrates this phenomenon using an uplink data transmission trace captured from the Amarisoft cell (Table 1). The top two subplots display the time series of allocated PRBs and the selected MCS for the UE. Starting from time 0.6 s (marked by ① in Fig. 12), the MCS drops to very low values. Notably, during this period, the allocated PRBs also decrease, even in the absence of cross traffic. This occurs because the base station's scheduler assigns fewer PRBs to a UE with poor channel conditions to improve transmission reliability and resource efficiency.

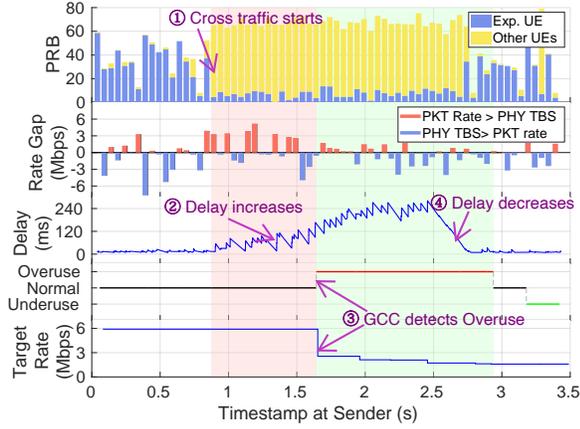

**Fig. 13— Cross traffic in 5G increases delay and degrades GCC target bitrate.**

The combined reduction in both MCS and allocated PRBs leads to a significant drop in the effective TBS. The third subplot shows the *rate gap*, calculated as the application's sending rate minus the estimated physical layer capacity (TBS converted to rate). During the low-MCS period (highlighted in red), the rate gap frequently becomes positive, indicating that the application is attempting to send data faster than the available physical layer capacity. This forces data to queue at the UE's RLC buffer. The fourth subplot confirms this, showing a corresponding increase in the RLC buffer size (②). As shown in the fifth subplot, this queue buildup directly results in increased packet one-way-delay (③), reaching values as high as 380 ms during this event. Subsequently, as the channel conditions improve (highlighted in green), the MCS and PRB allocation recover (④). The rate gap becomes negative, signifying that the physical layer capacity now exceeds the application's sending rate. This allows the accumulated buffer to drain, and the packet delay gradually decreases (⑤), returning to around 30 ms.

*5.1.2 Impact of cross traffic.* In cellular networks, the number of PRBs allocated to a specific UE is dependent on the demand from both itself and other UEs (i.e., cross traffic). When cross traffic increases, fewer PRBs are available for the test UE, impacting its achievable data rate.

Fig. 13 demonstrates this effect using a downlink trace from the T-Mobile 15 MHz TDD cell (Table 1). The first subplot shows the time series of PRBs allocated to our test UE and to other UEs. The presence of significant cross traffic, indicated by the yellow bars, commences at around time 0.8 s (marked by ①). During this period, the number of PRBs assigned to our test UE decreases substantially, leading to an immediate reduction in its physical layer capacity (TBS). The second subplot displays the rate gap: as TBS drops due to cross traffic, the application's sending rate exceeds available capacity, resulting in a positive rate gap and causing data to buffer at the UE, increasing latency (②).

Using logs from our instrumented WebRTC client, we also analyze the response of GCC. The fourth subplot shows GCC's network state estimation, while the fifth subplot displays its target sending bit rate. Approximately 0.8 s after the cross traffic begins, GCC detects an "overuse" state and reacts by reducing its target bit rate via multiplicative decrease (③). However, because the sending rate reduction is not instantaneous and initially remains above the constrained physical layer capacity, latency continues to increase, up to $\approx$ 250 ms. GCC continues to lower its sending rate until, around time 2.5 s, the application rate falls below the available 5G capacity. At this point, the UE's buffer begins to drain, and latency gradually decreases back towards the 30 ms level (④).

## 5.2 5G Protocol Timing and Reliability

Beyond the radio-resource variability discussed previously, the inherent operational timing and reliability mechanisms within the 5G protocols themselves also contribute significantly to packet delay and jitter. This section analyzes key protocol aspects, starting with uplink scheduling and followed by error recovery mechanisms (HARQ, RLC retransmissions), illustrating their impact on VCA performance with examples from our measurements.

*5.2.1 Uplink Scheduling.* As observed in our longitudinal experiments (§3), 5G often exhibits higher latency in the uplink compared to the downlink. This stems fundamentally from the uplink scheduling mechanism: Unlike in the downlink where the base station controls data transmission directly, for the uplink, the base station typically does not have a-priori knowledge of when or how much data a UE needs to send. Consequently, 5G employs a request-grant procedure for uplink resource allocation.

Time Division Duplexing (TDD) shares time slots between downlink and uplink, while Frequency Division Duplexing (FDD) uses separate frequency bands for simultaneous downlink and uplink transmission. The standard mechanism [1] to enable uplink transmission, illustrated conceptually in Fig. 15a (TDD) and Fig. 15b (FDD) operates as follows: When new data arrives in the UE's transmission buffer, the UE

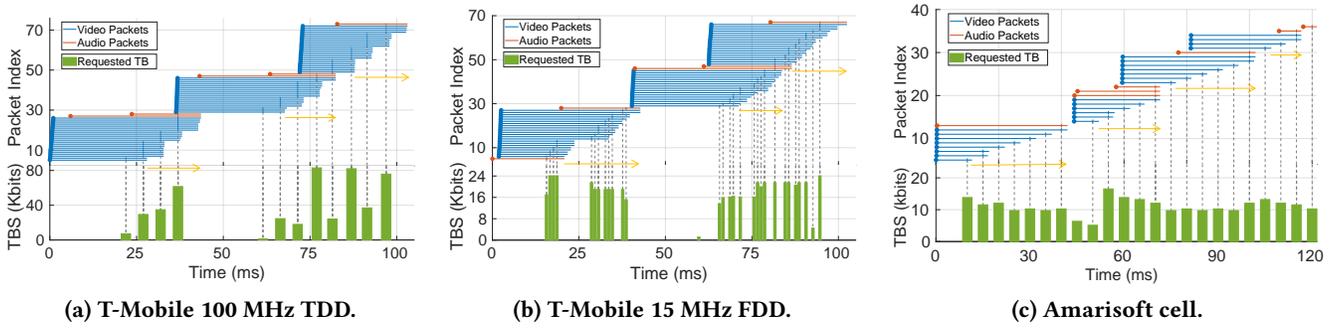

**(a) T-Mobile 100 MHz TDD.**   **(b) T-Mobile 15 MHz FDD.**   **(c) Amarisoft cell.**

**Fig. 14— Time series examples of WebRTC over 5G traces, combining transport-layer packet information with PHY-layer TBs. Dashed lines indicate the mapping between each packet and the corresponding TB that carries it.**

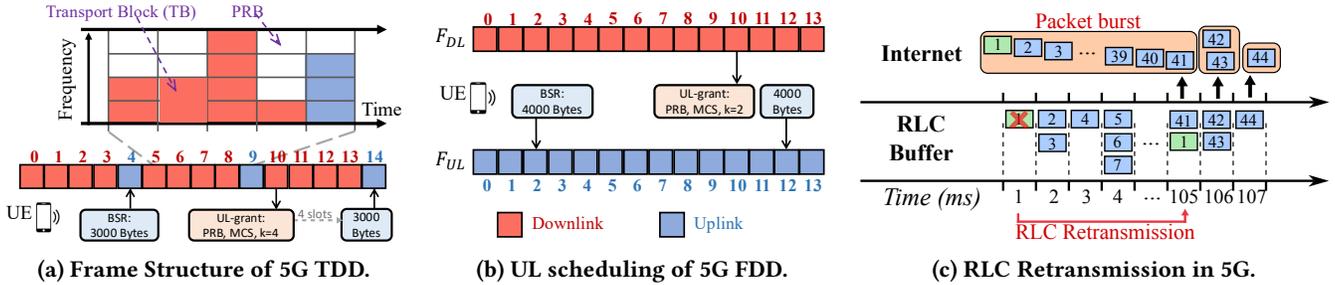

**(a) Frame Structure of 5G TDD.**   **(b) UL scheduling of 5G FDD.**   **(c) RLC Retransmission in 5G.**

**Fig. 15— 5G Data Transmission Fundamentals: (a) frame structure and UL scheduling of 5G TDD, (b) UL scheduling of 5G FDD, and (C) Head-of-Line blocking caused by RLC retransmissions.**

sends a Buffer Status Report (BSR) to the base station during the next available BSR opportunity, where the BSR indicates the amount of queued data. Upon receiving and processing the BSR, the base station allocates uplink resources (PRBs and MCS) to the UE via an uplink-grant message. The UE can then use these granted resources for transmission. However, a non-negligible scheduling delay exists between the UE sending the BSR and receiving/utilizing the corresponding grant [31]. In our measurements across the four 5G cells, this delay ranged from approximately 5 ms to 25 ms.

To analyze the practical impact of this scheduling delay on VCA traffic, Fig. 14 presents time-series traces from three distinct cells: a T-Mobile 15 MHz FDD cell, a T-Mobile 100 MHz TDD cell, and an Amarisoft cell. The upper portion of each figure visualizes individual packets as horizontal lines; the line's start and end point mark the sender's transmission time and receiver's reception time, respectively, thus the line length represents the one-way delay. The lower portion shows the corresponding physical-layer TBS over time.

VCAs typically generate data in bursts, where multiple packets constituting a single video frame are sent at once (see clustered transmit times in Fig. 14). Due to the size of these bursts and the limited TBS per grant, transmitting a full video frame often requires multiple consecutive Transport Blocks (TBs). As seen in Fig. 14b, the packets within a burst arrive spread out over time at the receiver. This intra-frame arrival-time variation is called *delay spread* [35] (indicated by yellow arrows in Fig. 14) and directly contributes to jitter.

Across the three cells, we observe different patterns of delay spread. In the T-Mobile 100 MHz TDD cell (Fig. 14a), the high bandwidth allows more packets to fit into a single TB, making the effect of delay spread less pronounced. The T-Mobile 15 MHz FDD cell (Fig. 14b) features frequent uplink-transmission opportunities with inherently smaller gaps between them due to its FDD pattern. However, its lower bandwidth (15 MHz) results in smaller TBS. This, in turn, necessitates transmitting more than 10 TBs per frame, leading to a large delay spread. Finally, in the Amarisoft cell (Fig. 14c), the uplink channel conditions are poor, leading to smaller TBS and lower uplink capacity. This forces the WebRTC sender to adapt to a lower bitrate (fewer packets per burst), but the effect persist.

To mitigate the UL scheduling delay, some 5G cells utilize proactive uplink grants. Our Mosolabs private 5G cell employs this strategy, as shown in Fig. 16. Proactive grants (blue bars) pre-allocate small amounts of resources before a BSR is received. These allow the first few packets of a burst to be sent earlier, reducing their latency (by approx. 10 ms in our trace). Once the BSR-triggered grant (green bars) arrives, the remaining buffered packets are transmitted.

While proactive scheduling reduces first-packet latency, it has drawbacks for bursty VCA traffic. First, it provides little benefit to the last packet's latency in a burst, thus barely improving the overall frame-level delay. Second, it can be inefficient; proactive grants may go unused if no data is ready (wasted bandwidth, visible as unfilled blue bars in Fig. 16). Third, it can lead to over-granting: the BSR reflects the buffer status when sent, but by the time the BSR-requested grant is usable, some data has already been delivered via proactive grants. This means the requested grant might be larger than necessary, potentially resulting in unused and wasted capacity within that grant (e.g., unfilled green bars in Fig. 16). Despite variations in configuration (FDD/TDD, bandwidth, proactive grants), the fundamental uplink scheduling process consistently introduced UL scheduling delay and delay spread for VCA traffic across all four 5G cells we measured.

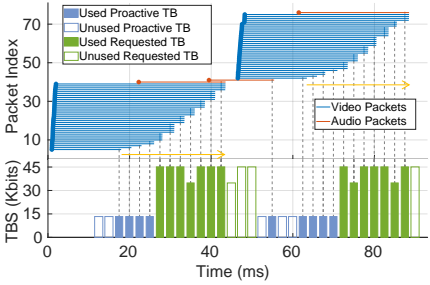

**Fig. 16— The Mosolabs cell employs proactive UL grants, leading to bandwidth waste (unfilled bars).**

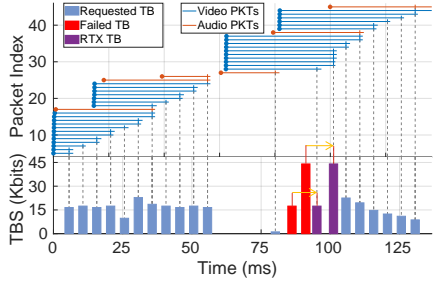

**Fig. 17— HARQ ReTXs inflate the packet delay by 10 ms, denoted as yellow arrows.**

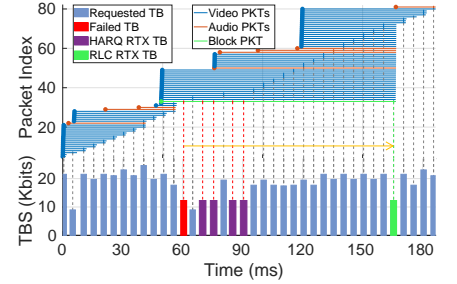

**Fig. 18— RLC ReTXs inflate the packet delay by 105 ms, denoted as the yellow arrow.**

*5.2.2 HARQ Retransmissions.* To combat the inherent susceptibility to errors of wireless networks, at the MAC layer, 5G employs Hybrid Automatic Repeat reQuest (HARQ). When the receiver fails to decode a TB, it signals a negative acknowledgment (NACK) back to the sender, prompting a HARQ retransmission of the same TB. While crucial for reliability, each HARQ retransmission attempt introduces additional delay for the packets contained within that TB.

Fig. 17 provides a time-series trace from our Amarisoft cell illustrating this process. TBs that initially failed decoding are highlighted in red, while subsequent successful HARQ retransmissions of those TBs are marked in purple. For packets carried in such retransmitted TBs, their one-way delay (the length of the horizontal packet lines) increases. As indicated by the yellow arrows in Fig. 17, each HARQ retransmission cycle under this cell adds 10 ms to the packet delay. If a retransmitted TB also fails, further HARQ attempts may occur (up to a configured limit), cumulatively increasing latency in multiples of this 10 ms delay and impacting overall latency.

HARQ retransmissions are relatively common, especially under challenging channel conditions or when the network employs aggressive MCS selection (prioritizing rate over robustness). Such events happen across all four measured cells in both downlink and uplink. In typical WebRTC sessions under our experimental setup, we observe hundreds of HARQ retransmissions per minute, although the exact frequency is highly dependent on the cell's specific physical layer rate control algorithm and the radio environment.

*5.2.3 RLC Retransmissions.* The 5G MAC layer imposes a configurable upper limit on the number of HARQ retransmissions for a single TB. In our Amarisoft cell's configuration, for instance, this limit was set to four attempts. If a TB remains undecodable after exhausting the four HARQ retries, the MAC layer abandons the transmission. Recovery responsibility then falls to the Radio Link Control (RLC) layer, situated above the MAC layer in the 5G protocol stack.

RLC retransmissions, however, incur substantially higher latency penalties. Fig. 18 shows a trace segment from the Amarisoft cell capturing an RLC retransmission event. Here, the initial failed TB (denoted as red bar) carries a specific packet (represented by the horizontal green line). Subsequent four HARQ retransmissions (denoted as purple bars) also failed. Consequently, the RLC layer initiated its own retransmission. The packet was finally delivered successfully via an RLC-retransmitted TB (denoted as green bar) that arrived approximately 105 ms after the initial failed transmission attempt, resulting in a delay inflation of 105 ms for this packet (indicated by the yellow arrow in Fig. 18). The precise delay incurred by RLC depends on various factors, including RLC protocol timers and base station configurations. Furthermore, RLC enforces in-order delivery of data segments to higher layers. This mandate creates a Head-of-Line (HoL) blocking problem when RLC retransmissions occur. As illustrated in Fig. 15c, packet 1 requires RLC retransmission, subsequent packets (2–41 in the example) that were successfully received at the MAC layer are held in the base station's RLC buffer. Once packet 1 arrives via RLC retransmission, the entire sequence of buffered packets (1–41) is released upwards nearly simultaneously. This HoL blocking explains the pattern observed in Fig. 18, where a large cluster of packets exhibit almost identical reception times (right edges of the lines),

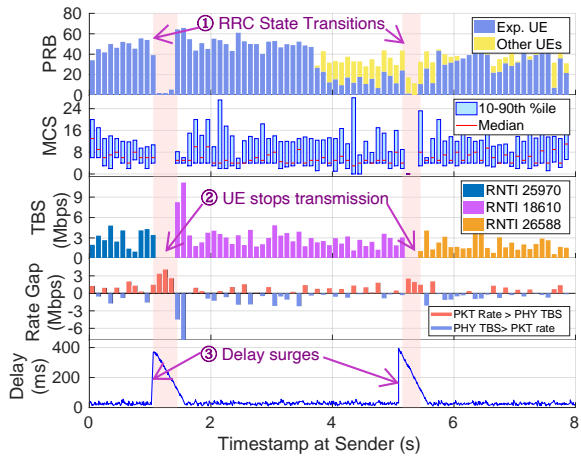

**Fig. 19— RRC state transitions halt PHY-layer transmissions, causing packet delay to spike to 400 ms**

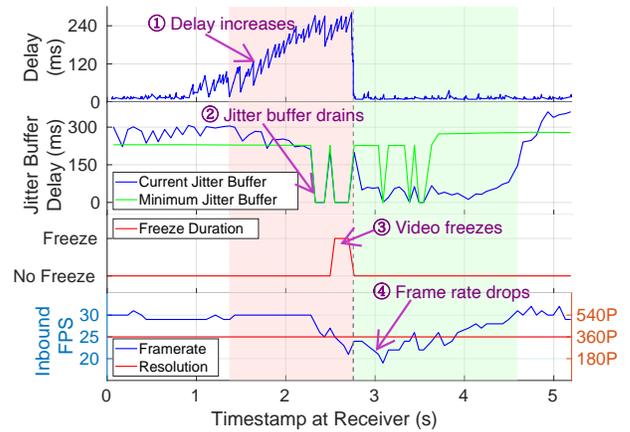

**Fig. 20— Rapid packet delay surges drain the jitter buffer, causing frame rate drops and video freezes.**

despite being transmitted over the air in different TBs earlier.

### 5.3 RRC State Transitions

Beyond the performance variations caused by radio resource dynamics and standard protocol timing/retransmissions, we observe a distinct and highly disruptive event during measurements on the T-Mobile 15 MHz FDD cell: unexpected Radio Resource Control (RRC) state transitions. Specifically, the UE undergoes RRC Release and subsequent RRC connection establishment while actively transmitting WebRTC traffic. This behavior deviates from the standard expectation that a UE remains in the RRC Connected state as long as active data transfer persists. While the precise triggers are unknown without access to gNB logs, potential causes include aggressive network inactivity timers, specific connection management policies, or transient Radio Link Failures.

Fig. 19 presents a data trace capturing such events. The first two subplots display the allocated PRBs and selected MCS over time. During the periods highlighted in red, both lack data points, indicating a complete cessation of PHY-layer transmissions. The third subplot confirms the underlying cause by showing a change in the Radio Network Temporary Identifier (RNTI). The RNTI is a MAC layer identifier assigned to a UE specifically when it is in the RRC Connected state; a change in RNTI signifies that the UE has transitioned out of and back into the RRC Connected state. This entire transition process, as measured in our traces, resulted in an interruption period of approximately 300 ms during which the UE could neither send nor receive data. Critically, the application layer (WebRTC) is unaware of this temporary network disconnection. The fourth subplot, showing rate gaps, indicates that the application continues to generate and send data throughout the interruption periods highlighted in red. Consequently, packets buffer extensively at the UE, causing one-way delay to surge dramatically (up to 400 ms), as shown in the fifth subplot.

This phenomenon of RRC transitions during active data transfer was unique to the T-Mobile 15 MHz FDD cell. Furthermore, its occurrence on this network was intermittent; sometimes the connection remained stable for hours, while at other times, these transitions occurred frequently, up to 3-4 times per minute.

## 6 Consequences of 5G Network Variability

This section outlines the consequences for VCA QoE, covering both the direct impact on media transmission and the indirect impact through congestion-control decisions.

### 6.1 Impact on Media Reception

Sudden delay increases and jitter as often caused by the cellular network, directly affect the playout quality of audio and video in video conferencing. VCAs use an adaptive jitter buffer to mitigate delay variance by temporarily buffering incoming video frames or audio samples before playback. To balance smooth media playback with low latency, the jitter buffer dynamically adjusts its size: it expands during poor network conditions and contracts when latency is stable [11].

Consequently, severe and rapid network delay fluctuations as seen in §5 increase the jitter buffer length leading to smooth playback but long end-to-end delay, which also compromises QoE. Fig. 20 illustrates such a scenario using a trace from the T-Mobile 15MHz FDD cell. Subplot 1 shows the one-way delay progressively increasing, reaching ≈280 ms (marked by ①). During this period, the network delay exceeds the jitter buffer's capability to compensate. The jitter buffer drains ② and the video freezes ③/④. The situation begins to improve at time ≈2.8 s (indicated by the gray vertical line), when the packet delay sharply decreases to ≈20 ms.

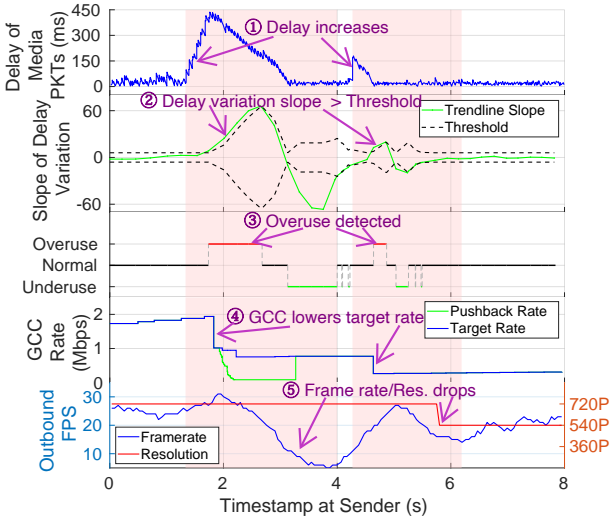

**Fig. 21— Increased media packet delay triggers GCC to lower the target bitrate, causing reduced frame rate.**

Following network recovery, the jitter buffer starts to rebuild (green highlighted area in subplot 2), and the video freeze ends. Nevertheless, the frame rate remains below 30 while the jitter buffer fills up again. Full recovery to the target frame rate of 30 fps (subplot 4) was achieved at time 4.6 s.

### 6.2 Impact on GCC Target-Rate Control

Beyond affecting actual media delivery, the performance variability in 5G networks impacts the rate (and consequently media quality) at which VCAs send. WebRTC uses GCC [7] to estimate link capacity and send rate based on network delay measurements. We now describe how 5G-induced delay fluctuations directly impact GCC's rate estimation, leading to lower-than-necessary send rates, hurting session quality.

**Mechanism.** GCC's sender-side rate control logic uses two main components: the delay-based estimator and the loss-based estimator. The delay-based estimator analyzes the one-way delay gradient as reported from receivers via RTCP. GCC applies a trendline filter to these readings to detect congestion build-up, ideally before significant packet loss occurs. This mechanism together with the loss-based estimator is used to estimate the available link capacity and compute a target send rate [16]. This rate is then further adjusted in a second step by the pushback controller (§6.3).

**Tracing Target-Rate Drops.** Fig. 21 illustrates the reaction of GCC's delay-based estimator to sudden delay increases induced by the 5G network. The figure highlights two distinct delay increase events and GCC's reactions. Subplot 1 displays the one-way packet delay, with the first highlighted region showing a surge to ≈440 ms ①. Subplot 2 presents the delay slope, a critical signal for congestion detection, obtained from GCC's internal state using our instrumented WebRTC client. A positive slope signifies increasing delay (potential congestion), while a negative slope suggests decreasing delay. GCC compares this slope against an adaptive threshold (gray dashed line in subplot 2) to classify the network state as "overuse," "underuse," or "normal" (subplot 3). During the first event, the positive delay slope leads GCC to detect an "overuse" state ③. Consequently, GCC sharply reduces its target rate via multiplicative decrease ④. This, in turn, causes a drop in the outbound video frame rate ⑤. Shortly thereafter, as the network delay begins to decrease, an "underuse" state is detected. In this state, GCC aims to stabilize the queue by maintaining the bitrate. Once the delay stabilizes and the state transitions to "normal," GCC resumes probing available bandwidth through additive increase of its target rate. The second delay event depicted in Fig. 21 triggers a similar cycle of detection and reaction, with a further bitrate reduction in this instance, leading to a decrease in video resolution from 720p to 540p. We also observe a noticeable delay between the onset of the network delay increase (subplot 1) and GCC's state change (subplot 3). This delay stems from the reliance on periodic RTCP feedback.

**Slow Rate Recovery.** 5G-induced delay can be short-lived at the network layer, for example when caused by transient cross traffic or temporary signal degradation. The application-level recovery by GCC, however, is often considerably slower. After an overuse event, GCC's default mechanism for probing available bandwidth uses cautious additive increase. Our measurements indicate that this additive increase phase can take over 30 seconds to restore the target bitrate to its pre-congestion level. The gap between rapid network-level recovery and slower application-level adaptation can lead to inefficient utilization of available radio resources and can prolong periods of degraded performance.

**GCC Acknowledged Bit Rate Estimator.** GCC incorporates a complementary mechanism known as the acknowledged bitrate estimator, which can facilitate faster recovery under certain conditions. This estimator calculates a bitrate based on the timestamps and sizes of packets acknowledged by the receiver, effectively measuring the actual throughput recently achieved. If a delay-based overuse event is short-lived while the acknowledged bitrate estimator concurrently reports sustained high throughput, WebRTC may prioritize this direct throughput measurement for rate control, effectively bypassing a prolonged slow additive increase and rapidly restoring send rate. In our experiments, we observed instances where this fast recovery mechanism successfully restored GCC's target bitrate to pre-congestion levels within ≈2 seconds, thereby minimizing the impact of short-lived overuse events. However, analysis with our anomaly detector (§4.2) reveals that such fast recovery occurs in only 1% of the detected anomalies. For the majority of events, GCC defaults to the slower additive increase process for recovery.

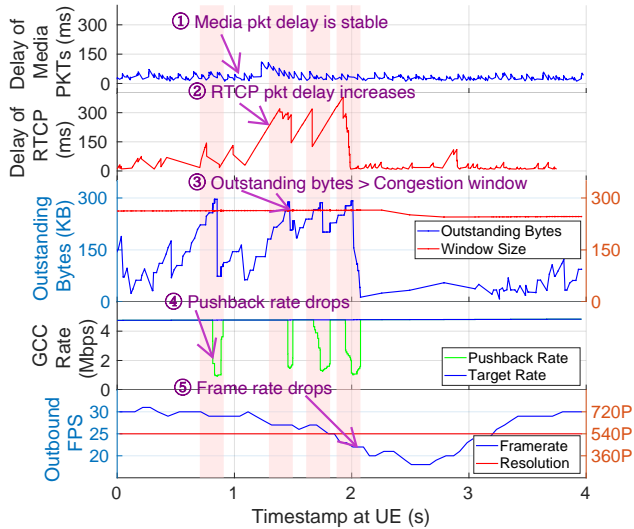

**Fig. 22—** RTCP packet delay causes outstanding bytes to exceed the congestion window, leading to a lower pushback rate and frame rate drop.

## 6.3 Impact on GCC Pushback-Rate Control

On top of the target bit rate (§6.2), the actual media send rate can be further constrained based on the amount of outstanding, unacknowledged bytes. We call the controller responsible for this final part of the send-rate calculation the *pushback rate controller*. This rate is ultimately provided to the video encoder and pacer [16].

**Mechanism.** GCC maintains a congestion window and tracks the volume of outstanding bytes. Under stable conditions with timely acknowledgments, the pushback rate aligns with the target bit rate. However, if outstanding bytes accumulate and exceed the congestion window limit, the pushback controller reduces the pushback rate. This reduction aims to prevent further network queuing and allow the volume of outstanding bytes to decrease as ACKs arrive, consequently impacting video frame rate or resolution. We provide more details on this mechanism in Appendix E.

**Tracing Pushback-Rate Drops.** The efficacy of this pushback rate control loop depends on the timely flow of media packets from sender to receiver and RTCP feedback from receiver to sender. A significant delay increase in either the forward (media) path or the reverse (RTCP feedback) path can cause the number of outstanding bytes to accumulate, potentially triggering the pushback rate drops. For instance, the substantial media packet delay increase (≈440 ms) shown in subplot 1 of Fig. 21 not only affects the bandwidth estimator but also contributes to the accumulation of outstanding bytes. This causes the pushback rate to diverge from the target bitrate estimated by the bandwidth controller, as illustrated in subplot 4 of that figure.

A clearer illustration of the pushback controller's distinct impact is presented in the data trace in Fig. 22. The forward media packet delay (① in subplot 1) remains stable throughout the observation period. Consequently, GCC's bandwidth estimator perceives no congestion, and the target bitrate (blue curve in subplot 4) remains stable and high. However, ② in subplot 2 reveals a significant increase in the delay of RTCP packets (reverse path), reaching over 300 ms in the highlighted red regions. This inflation of RTCP delay directly causes the number of outstanding bytes to accumulate (blue curve in subplot 3), eventually exceeding the congestion window limit (③). As a direct result of breaching this limit, the pushback controller intervenes, sharply reducing the pushback rate (④ in subplot 4) despite the stable target bitrate. This reduction in the effective sending rate leads to a corresponding drop in the transmitted video frame rate (⑤ in subplot 5). This example underscores how 5G-induced delays, even if confined to the feedback path, can trigger significant performance degradation at the application level through mechanisms like the pushback controller.

## 7 Related Work

**Video-Conferencing Measurement Studies.** The seminal work on the measurement of video conferencing applications (VCAs) by Baset et. al. [4] provided first insights into the operation of Skype. Since, studies have focused on the behavior and provided quality of VCAs in different environments [6, 9, 10, 23, 29, 32]. Others have focused on specific applications, analyzing their protocols, media formats, and congestion control [21, 24, 27]. A close related work by Yi *et al.* presents a first analysis of the behavior of Zoom in 5G [35]. In contrast, our work goes much further by identifying exact causal relationships between cellular network events and their impact on VCAs through a detailed analysis of the internals of both the RAN and WebRTC.

**Cellular-network measurement tools/studies.** There is a wide range of cellular-network measurement tools [20, 22, 33]. Out of these, we leverage NR-Scope [33]. Based on such RAN telemetry, systems have been developed to improve various aspects of applications, mostly focusing on congestion control [34]. Other works have studied the impact of cellular networks on a wide range of applications [13, 14, 19, 26, 30]. Finally, cross-layer measurements have been used to study the impact of 5G on the performance of video streaming [19]. Our work differs in that it focuses on RTC applications and provides a much more fine-grained analysis of the causal relationships between cellular network events and their impact on the application layer.

## 8 Conclusion

In this work, we present the first in-depth, cross-layer analysis of the performance of RTC applications in 5G. We identify a series of causal relationships between cellular-network events and their impact on RTC applications, which we model as a graph. Domino is a diagnosis tool that uses this graph to enable researchers, network operators, and application developers to understand and address performance issues in RTC applications in 5G cellular networks.

## Acknowledgements

This work has been supported by the National Science Foundation under Grant Nos. AST-2232457, CNS-2223556, and OAC-2429485.

## A  Ethics

The Zoom API data used in this study was anonymized with a one-way hash. All data exports were inspected and sanitized by a network operator to remove all personal data before being accessed by researchers. This study has been conducted with necessary approvals from our institution, including its Institutional Review Board (IRB).

## B  Video Resolution of WebRTC on 5G

Video resolution, presented in Table 3, reveals UL streams generally maintain higher resolutions than their DL counterparts.

| | T-Mobile TDD | | T-Mobile FDD | | Amarisoft | | Mosolabs | |
|---|---|---|---|---|---|---|---|---|
| 360p | 1.7% | 93.8% | 1.6% | 54.2% | 35.5% | 95.5% | 0.3% | 93.3% |
| 540p | 94.5% | 3.3% | 94.2% | 43.0% | 53.0% | 1.5% | 95.5% | 3.7% |
| 720p | 2.1% | 1.7% | 2.2% | 1.6% | 9.9% | 1.8% | 2.2% | 1.7% |
| 1080p | 1.7% | 1.2% | 2.0% | 1.2% | 1.6% | 1.2% | 2.0% | 1.3% |

**Table 3**— Video resolution distribution of UL (**blue**) and DL (**red**) streams stratified by 5G cell type.

## C  Statistics of all Causal Chains

Here we list the causality chain happening ratio when a consequence happens in the WebRTC application in Table 4. We only count one when a consequence is caused by multiple causes in the 5G network, which is why the total number in the table doesn't add up to 100 %.

## D  Complete Event Detection Conditions

Here we list all the event detection conditions discussed in Fig. 9 in Table 5. For the application events (1 to 10 in Table 5), we extract the features from both local and remote clients. For the bi-directional 5G network events (13 to 18), we extract the features from both UL and DL 5G resource allocation logs from NR-Scope [33] or gNB log. Adding them all up, there are $2 \times 10 + 6 \times 2 + 4 = 36$ dimensions in the feature vector. Here we provide the detailed calculation of each of them.

**Inbound/Outbound frame rate** ↓ **(1, 2):** We first compare the maximum and minimum frame rates with the threshold, then we ensure that the minimum value happens after the maximum value.

$$\max(\text{frame\_rate}) > 27, \text{ and } \min(\text{frame\_rate}) < 25, \text{ and}$$
$$\arg\max_i \text{frame\_rate} < \arg\min_j \text{frame\_rate}.$$

**Outbound resolution** ↓ **(3):** Since the resolution drops from one step to another (*e.g.* 540P to 360P), as long as there is a resolution drop, we set the resolution drop to be true for this window.

$$\exists i \in [0, W-1], \text{resolution}[i+1] < \text{resolution}[i].$$

**Jitter buffer drain** ↓ **(4):** If there is a point where the jitter buffer delay is 0 milliseconds.

$$\exists i \in [0, W], \text{jitter\_buffer\_delay}[i] = 0ms.$$

**Target bitrate** ↓ **(5):** Normally, the target bitrate stays the same or increases as the GCC probes up when nothing bad happens. As long as there is a target bitrate drop, it means something bad happens.

$$\exists i \in [0, W], \text{target\_bitrate}[i+1] < \text{target\_bitrate}[i].$$

**GCC overuse detected (6):** Whenever there is an overuse entry in the GCC logs within the window.

$$\exists i \in [0, W], \text{gcc\_netstate}[i] = \text{'overuse'}.$$

**Pushback rate** ↓ **(7):** Normally, the pushback rate aligns with the target bitrate and has an identical non-decreasing trend.

$$\exists i \in [0, W-1], \text{pushback\_rate}[i+1] < \text{pushback\_rate}[i].$$

**Congestion window full (8):** The outstanding bytes are the inflight bytes, and the window occupancy is the ratio between the outstanding bytes and the window size. We first calculate the congestion window ratio by performing the element-wise division between the outstanding bytes and the GCC congestion window. Then, we check if there is a point where the window ratio exceeds 1.

$$\text{cwnd\_ratio} = \text{outstanding\_bytes/gcc\_cwnd};$$
$$\exists i \in [0, W], \text{cwnd\_ratio}[i] > 1.$$

**Outstanding bytes** ↑ **(9):** We use a different small window (indexed by $k$) of 10 samples to calculate the average outstanding bytes and detect whether there is an uptrend.

$$\text{wind\_bytes}[k] = \frac{1}{10}\sum_{i=10k}^{10(k+1)} \text{outstanding\_bytes}[i],$$
$$k = 0, ..., W/10 - 1,$$
$$\exists k \in [0, W/10-1], \text{wind\_bytes}[k+1] > \text{wind\_bytes}[k].$$

**Pushback rate is unequal to target bitrate (10):** We detect if there is any sample where the target bitrate is unequal to the pushback rate.

$$\exists i \in [0, W], \text{target\_bitrate}[i] \neq \text{pushback\_rate}[i].$$

**Forward/reverse packet delay** ↑ **(11, 12):** Similarly, we maintain windows for the packet delay (indexed by $k$) and use the average delay as the trend. Then we detect whether

| | Poor Channel | | Cross Traffic | | UL Scheduling | | HARQ ReTX | | RLC ReTX | | RRC State | | RLC ReTX | |
|---|---|---|---|---|---|---|---|---|---|---|---|---|---|---|
| **Jitter Buffer Drains** | 30% | (0%) | 2.7% | (0%) | 2.4% | (0.45%) | 4.6% | (0.45%) | 0% | (0.22%) | 0% | (0%) | 0% | (0.45%) |
| **Target Bitrate ↓** | 43% | (0.90%) | 40% | (0.00%) | 16% | (14%) | 56% | (13%) | 0% | (0.45%) | 0.81% | (0%) | 0.27% | (34%) |
| **Pushback Rate ↓** | 56% | (2.3%) | 42% | (0%) | 34% | (45%) | 67% | (57%) | 0% | (0.45%) | 9.8% | (0%) | 1.4% | (4.7%) |

**Table 4— Each causal chain's ratio over all detected chains, numbers in the brackets are from the private 5G cells.**

| Events/Features | Condition |
|---|---|
| 1. Inbound frame rate ↓ | Maximum inbound frame rate is higher than 27, while the minimum inbound frame rate is smaller than 25. |
| 2. Outbound frame rate ↓ | Maximum outbound frame rate is higher than 27, while the minimum outbound frame rate is smaller than 25. |
| 3. Outbound resolution ↓ | There is a downtrend in outbound resolution. |
| 4. Jitter buffer drains | The client's jitter buffer drops to 0 milliseconds. |
| 5. Target bitrate ↓ | There is a downtrend in the client's target bitrate. |
| 6. GCC overuse detected | There is an overuse entry in the GCC log. |
| 7. Pushback rate ↓ | There is a downtrend in the client's pushback rate. |
| 8. Congestion window full | The client's outstanding bytes are bigger than the client's GCC congestion window bytes. |
| 9. Outstanding bytes ↑ | There is an uptrend in the client's windowed outstanding bytes. |
| 10. Pushback rate unequal to target bitrate | If these two values are not equal to each other at any point. |
| 11. Forward packet delay ↑ | There is an uptrend in the windowed forward packet delay. |
| 12. Reverse packet delay ↑ | There is an uptrend in the windowed reverse packet delay. |
| 13. Allocated TBS ↓ | Minimum TBS is smaller than 80 percent of the maximum TBS in the window. |
| 14. App bitrate exceeds the allocated TBS | The percentage of time when App bitrate exceeds the allocated TBS is higher than 10 %. |
| 15. 5G cross traffic | Other UE's allocated PRB summation is higher than 20 % of our UE's allocated PRB summation. |
| 16. 5G channel degrades | Maximum 90-percentile of the grouped MCSs (with a window of 50 ms) is smaller than 20, and the less-than-10 medium value appears more than 10 times. |
| 17. HARQ retransmission | There are more than 20 instances of HARQ retransmission detected. |
| 18. RLC retransmission | The gNB's log indicates RLC retransmission. |
| 19. Uplink scheduling delay | As long as the transmission uses the 5G uplink channel. |
| 20. RRC state change | The UE's RNTI changes during the window. |

**Table 5— Event detection conditions used by Domino for feature extraction in the sliding window.**

there is an uptrend in the windowed packet delay and the maximum delay is higher than 80 ms.

$$\text{wind\_delay}[k] = \frac{1}{10}\sum_{i=10k}^{10(k+1)} \text{packet\_delay}[i],$$

$$\exists k \in [0, W/10 - 1], \text{wind\_delay}[k+1] > \text{wind\_delay}[k], \text{ and}$$

$$\exists i \in [0, W], \text{packet}_d\text{elay}[i] > 80ms.$$

**Allocated TBS ↓ (13):** We determine the allocated TBS drop if the min TBS drops to below 80 % of the max TBS.

$$0.8\max(\text{TBS}) > \min(\text{TBS}),$$

$$\arg\max_i \text{TBS} < \arg\min_i \text{TBS}.$$

**App bitrate exceeds the allocated TBS (14):** The amount of time when App's bitrate exceeds the bitrate calculated from TBS exceeds 10 % of the total sliding window time.

$$\text{rate\_diff} = \text{app\_bitrate} - \text{tbs\_bitrate},$$

$$\text{sum}(\text{rate\_diff} > 0) > 0.1W.$$

**5G cross traffic (15):** When the PRB allocated for other UE ($j = 1 : n$) exceeds 20 % of our UE.

$$\sum_{i=0}^{W}\sum_{j=1}^{n} \text{PRB}_j[i] > 0.2\sum_{i=0}^{W} \text{PRB}_0[i];$$

**5G channel degrades (16):** We first group the MCS of our

UE with a 50 ms window, and if the grouped lower than 10 MCS groups' 50-percentile appears more than 10 times and the 90-percentile is lower then 20.

$$\text{wind\_MCS}[k] = \frac{1}{10}\sum_{i=k*50ms}^{(k+1)*50ms} \text{MCS}[i], k \in [0, W/0.05 - 1]$$

$$\text{wind\_MCS}_{90th} < 20 \text{ and sum}(\text{wind\_MCS}_{90th}),$$

$$\text{sum}(\text{MCS}_{50th} < 10) > 10.$$

**HARQ retransmission (17):** The HARQ retransmission happens constantly, but only certain severe cases cause the performance impairment. So Domino regards the HARQ retransmissions when there are more than 10 HARQ retransmissions.

$$\sum_{i=n}^{W} \text{HARQ\_retx}[i] > 10.$$

**RLC retransmission (18):** We determine the RLC retransmission when the gNB log from Amarisoft shows an RLC retransmission entry.

**Uplink Scheduling (19):** Whenever there is an uplink channel transmission, we set this feature to be true.

**RRC state change (20):** We check our client's RNTI and report this feature when there is a change.

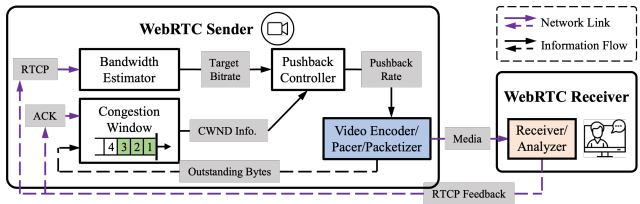

**Fig. 23— GCC combines bandwidth estimation and congestion window size to compute the pushback rate.**

## E    WebRTC Pushback-Rate Mechanism

The GCC pushback controller takes both target rate and congestion window information into account, and outputs a pushback rate. The congestion window information includes both the congestion window size and the outstanding bytes, and the valumn of outstanding bytes is affected by both forward path and reverse path network delays. This mechanism is illustrated in Fig. 23.